# Intragranular nucleation of tetrahedral precipitates and discontinuous precipitation in Cu-5wt%Ag

*M. Bonvalet[#,\*], X. Sauvage, D. Blavette*

Normandie Univ, UNIROUEN, INSA Rouen, CNRS, Groupe de Physique des Matériaux, 76000 Rouen, France

# Present affiliation: KTH Royal Institute of Technology, Department of Materials Science and Engineering, Brinellvägen 23, SE-100 44 Stockholm, Sweden

\* Corresponding author: bonvalet.manon@gmail.com

## Abstract

Both continuous and discontinuous precipitation is known to occur in CuAg alloys. The precipitation of Ag-rich phase has been experimentally investigated by atom probe tomography and transmission electron microscopy after ageing treatment of Cu-5%wtAg at 440°C during 30'. Both continuously and discontinuously formed precipitates have been observed. The precipitates located inside the grains exhibit two different faceted shapes: tetrahedral and platelet-shaped precipitates. Dislocations accommodating the high misfit at the interface between the two phases have also been evidenced. Based on these experimental observations, we examine the thermodynamic effect of these dislocations on the nucleation barrier and show that the peculiar shapes are due to the interfacial anisotropy. The appropriate number of misfit dislocations relaxes the elastic stress and lead to energetically favorable precipitates. However, due to the large misfit between the parent and precipitate phases, discontinuous precipitation that is often reported for CuAg alloys can be a lower energetic path to transform the supersaturated solid solution. We suggest that the presence of vacancy clusters may assist intragranular nucleation and decrease the continuous nucleation barrier. We eventually propose qualitative thermodynamic and kinetic justifications accounting for the relative importance of homogeneous and discontinuous precipitation modes as a function of temperature.

# 1. Introduction

The Cu-Ag binary system is of great technological and practical interest for the design of alloys dedicated to electrical industry. Depending on the Ag content, the thermal and mechanical treatments, different microstructures are generated. Ultra-low levels of silver (less than 0.1wt.%) are commonly used to control the grain size or to improve the thermal stability of drawn wires [1]. The low solubility of Ag in the fcc copper (less than 0.15wt.% at room temperature[2,3]) and the relatively large atomic size mismatch (about 12% [4]), lead indeed to a strong interaction with crystalline defects and especially boundaries. Besides, Cu-Ag alloys with silver contents ranging from 5 to 10 wt.% are also used when a combination of high strength (up to 1 GPa) and good electrical conductivity (up to 80% IACS) is required [5–7]. At such silver concentrations, a homogeneous solid solution can be obtained by annealing near 800°C [8] followed by rapid quenching. Precipitation treatment is then carried out at lower temperature (typically 400 to 500°C). The decomposition of metastable Cu-Ag solid solutions may proceed whether by continuous precipitation or discontinuous precipitation (also called cellular precipitation) [9,10]. During discontinuous precipitation (DP), nucleation of equilibrium phase starts at grain boundaries and then a transformation front progressively moves inward the grain forming a typical lamellar structure with specific orientation relationships between the fcc Cu and the fcc Ag phases [11]. Even if this lamellar structure can be refined by drawing to increase the mechanical strength of the alloy [5–7,12], continuous precipitation may also directly provide a significant strengthening if a fine distribution of nanoscaled Ag particles is achieved. It was shown that electrical and mechanical properties may be tailored by controlling the volume fraction, number density and size distribution of continuous precipitates [7]. It is then of major interest to handle these microstructure features and to identify precisely the precipitation mode according to the chemical composition and precipitation treatment to minimize DP or even suppress this precipitation mode.

DP in Cu-Ag alloys is favored by the large misfit between Cu and Ag phases (lattice mismatch is about 12%). The high elastic stress leads to a large barrier for homogeneous nucleation of coherent Ag nuclei in the FCC Cu-rich parent phase. It has been shown however that alloying with Zr or annealing in a well-defined temperature range promote the homogeneous precipitation of Ag [6,7,13,14]. The observed alignment of the precipitates along {111} habit planes was suggested to be related to the growth of stacking fault [15] or to the minimization of the elastic energy [14]. However, the exact underlying mechanisms are not clarified yet and to the knowledge of the authors no combination of Atom Probe Tomography (APT) and high resolution Scanning Transmission Electron Microscopy (STEM) have been reported in the literature to investigate homogeneous precipitates. The growth mechanism by atomic ledge migration, has also been reported [14]. However no information on the thermodynamic stability of such peculiar shape was given. It is interesting to note that in some cases, the homogeneous nucleation barrier may be reduced significantly if the solute content in the precipitates is smaller than the equilibrium concentration [16] since this reduces both the interfacial and the elastic energies. Elastic or interfacial energy anisotropies are also known to affect the morphology of "continuous" precipitates in metallic alloys that affects the mechanical properties as well. The continuous nucleation barrier can be decreased when nucleation starts at crystal defects such as dislocations or vacancy clusters that promote heterogeneous precipitation.

In this work we have investigated homogeneously nucleated precipitates in a Cu-5%wtAg alloy. Thanks to a combination of APT and high resolution STEM a special emphasis has been given on their composition and morphology to understand the potential thermodynamic stability of homogeneous nuclei and the mechanisms of growth of Ag in Cu. Both tetrahedral and plate-like precipitates were evidenced. Based on these observations and on thermodynamic aspects, the accommodation of the lattice at the interface between the matrix and the precipitates as well as the anisotropy of the interfacial energy are investigated, and the peculiar shape explained with a new thermodynamic model.

## 2. Experimental

The binary alloy investigated was delivered by Goodfellow with a nominal composition of Cu-5wt.%Ag. A homogenization treatment was carried out at 800°C during 2h followed by water quenching. Then, the precipitation treatment was conducted at 440°C during 30min to achieve phase separation through the nucleation and growth of fcc Ag precipitates in the CuAg parent phase. Microstructures were characterized by STEM using a JEOL-ARM 200F probe corrected microscope operated at 200kV. Images were recorded with a probe size of 0.1nm in both Bright Field (BF) and High Angle Annular Dark Field (HAADF). STEM samples were prepared by electropolishing (30% $HNO_3$ + 70%Methanol at -30°C and 12V) followed by short time ion milling (5 keV, ± 3°, 5min). Silver particles were also analyzed by APT with a LEAP 4000-HR (CAMECA). Samples were field evaporated at 50K using electric pulses (20% pulse fraction, repletion rate of 200kHz). Three dimensional reconstructions were computed using IVAS® software and further data processing was carried out with the GPM 3dsoftware. APT samples were prepared by conventional electropolishing methods.

## 3. Results

After aging at 440°C during 30min, only a relatively small amount of grains transformed by the DP mechanism was observed. One small grain transformed by this mechanism clearly appears on the HAADF image of Fig. 1 (a). It distinctly exhibits some typical lamella shaped Ag particles elongated along the growth direction (brightly imaged since the atomic number of Ag is significantly higher than that of Cu). The HAADF line profile computed across the transformation front (direction arrowed on Fig. 1 (a)) confirms that the Ag content is lower in the fcc Cu (far left of the profile) than in the untransformed fcc supersaturated solid solution (far right of the profile). Besides, it is interesting to note that there is a significant Ag enrichment at the transformation front (maximum of the profile) confirming that this boundary play a key role in the growth mechanism by promoting the fast redistribution of atoms between the fcc Cu rich and the fcc Ag rich phases. The DP process always starts at GBs but not along every GBs. However, as shown on Fig. 2 (a), these crystal defects are favorable nucleation sites for Ag particles and most of them are covered with particles that have grown up to a mean diameter in a range of 10 to 25nm during aging at 440°C. Dislocations are other typical crystalline defects. They are darkly imaged in the STEM-BF image of Fig. 2 (b) thanks to diffraction contrast. A high density of nanoscaled dark particles is also clearly exhibited. It is thought that these are Ag particles that have homogeneously nucleated in the supersaturated solid solution. However, many particles appear along the dislocation lines, indicating that like GBs, dislocations promote the heterogeneous nucleation of the high misfit fcc Ag phase.

A closer observation at high magnification of homogeneously nucleated Ag particles reveals that they exhibit a cube on cube orientation relationship with the fcc matrix (Fig. 3 (a)). Some moiré fringes appear on the image with a periodicity along the <111> direction of 1.9±0.1 nm, corresponding respectively to nine (111)Cu planes ($d_{(111)Cu}$ = 0.209) and eight (111)Ag planes ($d_{(111)Ag}$ = 0.236). Thus, these moirés are simply the result of the expected 12% lattice parameter mismatch between the fcc silver and copper phases. Besides, the filtered image displayed on Fig. 3(b) clearly reveals a series of (111) edge dislocations that relax the misfit and elastic energy at the interface between both phases (Fig. 3 (b)). When imaged along a (112) zone axis (Fig. 3 (a)) Ag particles often exhibit a peculiar triangle shape with a mean edge length ranging from 10 to 20 nm. Such a morphology corresponds to the projection of a tetrahedron with (111) facets. The three dimensional morphology of such particles can be better visualized in volumes collected by APT analyses as shown in Fig. 4 (a). It is well known that three-dimensional reconstructions may be affected by local magnification effects resulting from the different evaporation fields of precipitates and matrix [17]. The profile showing the evolution of the relative density across the particles (Fig. 4 (b)) indicates indeed that the evaporation field of copper is lower than that of silver but with a relatively small difference (Ag gives rise to a higher local magnification and hence to a lower relative density in the particle image). This certainly affects in some extent the size and the composition gradient across interfaces but its importance is small (small difference in density, Fig. 4 (b)). So the general morphology of the tetrahedron is thought to be preserved. The concentration profile (Fig. 4 (b)) indicates that the silver concentration inside precipitates lies in a range of 95 to 100at.%. This value is in good agreement with predictions computed using the Thermocalc® software [18] (98.2 at.% Ag). Consequently one may conclude that precipitates have reached their equilibrium concentration.

Next to tetrahedron precipitates, few plate shaped precipitates were also observed (Fig. 5 (a) (labelled P)). Such a particle has also been successfully analyzed within the matrix by APT (Fig. 5 (b)). To reveal completely the three dimensional morphology, the data have been filtered with a silver threshold of 20at.% and the volume rotated in two positions tilted by 90° on Fig. 5 (c). Besides, one should note that the local concentration, as measured thanks to concentration profiles, was very similar to that of tetrahedral precipitates (data not shown here).

APT analyses also provided the matrix composition in areas where homogeneous precipitation occurred (i.e. in most of the alloy excepted the few untransformed regions or the small grains transformed by DP such as on Fig. 1 (a)): for all analyses it stands in the range of 2.1 to 2.4 at.% Ag. This is significantly larger than the equilibrium solubility (about 1at.% at 440°C [2,8]) but also significantly lower than the concentration measured by APT after quenching and prior to aging (2.8±0.1 at.%) showing that the Ag supersaturation has decreased as a consequence of precipitation. The volume fraction of Ag precipitates, close to the molar fraction, after aging as derived from the measured phase composition and the so-called lever rule was thus estimated in range of 0.3 to 0.8% whereas the equilibrium volume fraction given by ThermoCalc and the TCCU2 database [18,19] is close to 4%.

This uncommon zoology of precipitates observed raises several open questions related to nucleation mechanisms. In particular, the faceted shapes underline the importance of elastic or/and interfacial anisotropies.

## 4. Discussion

Both TEM and APT investigations (Fig. 3-4) revealed that most of homogeneously nucleated precipitates displayed a tetrahedral shape with (111) orientated interfaces, (referred as (111) tetrahedron in the following). Some platelet (111)-shaped precipitates were also observed (Fig. 5). In the following, thermodynamic arguments are considered to understand these peculiar shapes. More particularly, the effect of dislocations accommodating the lattice mismatch, called thereafter misfit dislocations, on the homogeneous nucleation barrier is considered. One should note however that no particular emphasis is given on the competition between continuous and discontinuous precipitation since it does not imply only thermodynamic arguments, but also the complex interaction with kinetic effects, grain boundary structures and composition [20].

### 4.1. Morphologies of the precipitates

Since the Cu-Ag system exhibits a large misfit (12%) one expects a huge effect of the elastic contribution on the shape of precipitates. The dominance of the elastic contribution over the interfacial energy can be evaluated by calculating their ratio. The interfacial ($E_{int}$) and elastic ($E_{elas}$) energies of a tetrahedral incompressible precipitate of edge length $L$, surface $S$, volume $V$ and interfacial energy $\gamma$ are respectively:

$$E_{int} = \gamma S = \sqrt{3}\gamma L^2 \qquad \text{Eq. 1}$$

$$E_{elas} = \Delta G_{elas}.V = E\varepsilon^2 L^3 / \left(6\sqrt{2}(1-\upsilon)\right) \qquad \text{Eq. 2}$$

with $E$ the elasticity modulus, $\varepsilon$ the misfit and $\upsilon$ the Poisson ratio. Besides, the temperature dependence of the elasticity modulus writes as [21]:

$$E(T) = E(0K)(1 - 0.5T/T_m) \qquad \text{Eq. 3}$$

where $T_m$ is the melting temperature of Cu.

It is worth mentioning that Eq. 2 was initially derived by Eshelby [22] for a spherical particle. However the prefactor $\Delta G_{elas}$ represents an energy density and when the elastic energy is considered isotropic, it could be apply to any volume. In our case the volume of a tetrahedron.

Using the parameters listed in Tab. 1, the ratio $E_{elas}/E_{int}$ was found equal to 1 for $L = 1.44$ nm indicating that elastic effects dominate for $L > 1.44$ nm.

The faceted shape of the precipitates may be accounted by the anisotropy of elastic constants $C_{11}$, $C_{12}$ and $C_{44}$. Cahn [23] introduced a factor $\xi$ so as to predict the "soft elastic" direction as a function of elastic constants:

$$\xi = \frac{C_{11} - C_{12} - 2C_{44}}{C_{44}}$$

with $C_{11}$ = 166.1 GPa and $C_{12}$ = 119.9 GPa for Cu [24].

This factor being negative for Cu and Ag, it indicates that the "soft elastic" direction is <100>. Elastic anisotropies cannot therefore explain the tetrahedral (111) shapes observed.

Interfacial energies estimated by Monte Carlo simulations [11] ($\gamma_{111}$=230 mJ/m², $\gamma_{110}$=550 mJ/m² and $\gamma_{100}$=530 mJ/m²) indicate that they are strongly anisotropic and that the (111) interface has the lowest interfacial energy. It is important to note that a combination of (111) facets may precisely give rise to the tetrahedrons that have been experimentally observed in the present study. This suggests that the elastic contribution is relaxed in some way so that precipitate shape is mainly governed by the interfacial contribution. As discussed later and already shown experimentally, this relaxation is promoted by misfit dislocations (Fig. 3 (b)).

Let us consider the nucleation of a tetrahedral precipitate at constant pressure and temperature. The work of formation of a nucleus of edge length $L$ is given by:

$$\Delta G = (\Delta G_{ch} + \Delta G_{elas})V + \gamma \cdot S = (\Delta G_{ch} + \Delta G_{elas})L^3/6\sqrt{2} + \gamma \cdot L^2 \sqrt{3} \qquad \text{Eq. 4}$$

where $\Delta G_{ch}$ is the driving force for nucleation.

This work of formation is plotted on Fig. 6 (in black) as a function of the nucleus size $L$ at T=440°C. The interfacial contribution (grey dotted line), the elastic contribution (grey dashed line) and the chemical driving force (grey dashed-dotted line) are also represented. Parameters used are listed in table 1. The driving force for nucleation was calculated with the ThermoCalc software and the TCCU2 database [18,24]. Fig. 6 shows that the elastic term when not relaxed by dislocations is larger than the interfacial contribution. Besides, the work of formation is observed to constantly increase with the precipitate size and it never reaches a maximum. This indicates that homogeneous nucleation and growth are in principle impossible since the nucleation barrier is infinite. Therefore the precipitates observed experimentally have to nucleate through a mechanism for which the elastic stress is relaxed by dislocations.

### 4.2. Misfit dislocations

Dislocations were observed along the precipitate/matrix interfaces (see Fig. 3 (b)). We therefore considered edge dislocations located along the matrix/precipitates interfaces accommodating the misfit in our thermodynamic calculations. Their energy can be expressed as [25]:

$$E_{dislo} = \frac{\mu b^2 L_{dislo} \ln(R/r_0)}{4\pi(1-\upsilon)} \qquad \text{Eq. 5}$$

with $\mu$ the shear modulus ($\mu = E/(2(1+\upsilon))$), $b$ the Burger's vector ($b = d_{(111)_{Cu}} = a/\sqrt{3}$, where $a$ is the lattice parameter), $L_{dislo}$ the total dislocation length (sum on all the dislocation lines), $R$ the extension of the stress field (approximated by the average half distance between dislocations and derived below), and $r_0$ the core radius of the dislocation that we considered equal to the norm of the Burger's vector. $L_{dislo}$ and $R$ that are key parameters in Eq. 5 that will be derived thereafter.

We have made the assumption that the observed misfit dislocations are periodically distributed at the precipitate/matrix interface. In order to relax the lattice misfit for the four variants of (111) planes, four dislocations circuits are necessary. For the sake of clarity, only one variant of dislocation (Burgers' vector 1/3 <-1-11>) has been represented in Fig. 7. Each dislocation is a loop that runs all around the precipitate on three facets. Given the precipitate size shown in Fig. 7, three dislocations circuits are required to relax the misfit stress in this example. So as to relax the misfit for the 4 variants of (111) planes, one needs four variants of dislocation lines of Burgers vectors 1/3<111> 1/3<-111>, 1/3<-1-11>, 1/3<1-11> respectively. In Fig. 7, only one series of three 1/3 <-1-11> dislocations has been represented.

Since the four dislocation circuits are equivalent, the total dislocation energy is simply four times that of one dislocation variant. We have therefore introduced the number of dislocations of given Burger's vector type (N), i.e. one of the variant (in Fig. 7, N=3).

These misfit dislocations change the elastic energy of nuclei. The free enthalpy of formation of nucleates now writes:

$$\Delta G = (\Delta G_{ch} + \Delta G'_{elas})L^3/6\sqrt{2} + \gamma \cdot L^2 \sqrt{3} + E_{dislo} \qquad \text{Eq. 6}$$

where the new elastic contribution $\Delta G'_{elas}$ can be expressed as a function of the residual misfit $\varepsilon_{res}$:

$$\Delta G'_{elas} = E\varepsilon_{res}^2/(1-\upsilon) \qquad \text{Eq. 7}$$

that can be expressed as a function of $N$ because we consider a periodic distribution of the dislocations:

$$\varepsilon_{res} = \varepsilon - \frac{N}{N_{111}} \qquad \text{Eq. 8}$$

with $N_{111}$ the number of (111) planes per face. Eq. 8 reports the reduction of stored elastic energy when introducing $N$ dislocations to relax the lattice.

The two key parameters are the number of dislocations ($N$) and their total length ($L_{dislo}$). Introducing dislocations may decrease the elastic energy. However, this also has an energy cost (Eq. 5). A balance has therefore to be found.

The length of a dislocation line covering one facet can be expressed as a function of both the number of dislocations of the same Burger's vector variant per facet ($N$) and the edge length ($L$) of the tetrahedron:

$$L_i = \frac{i}{N+1}L \qquad \text{Eq. 9}$$

Since a similar number of dislocations per facet is considered, the total length of dislocations that is an input data in Eq. 5, i.e. the sum on all the dislocation lines, writes as:

$$L_{dislo} = 4\times 3 \times \sum_{i=1}^{N} L_i = 6NL \qquad \text{Eq. 10}$$

In Eq. 10, 4 represents the number of facets and 3 represents the number of dislocations with a different Burger's vector variant.

The extension of the stress field needed to calculate the dislocation energy (Eq. 5) is given by:

$$R = \frac{d_{dislo}}{2} = \frac{L}{\sqrt{6}(N+1)}$$  Eq. 11

with $d_{dislo}$ the distance between (111) dislocations on a facet.

Combination of Eqs. 5, 10 and 11 allows to evaluate the dislocation energy for a specific precipitate size and number of dislocations per facet. Let us now calculate $N_{111} = h/d_{111} = \sqrt{2}L/a$, with $h$ the height of the tetrahedral-shaped precipitate, and replace it in Eq. 8. Then the residual strain writes:

$$\varepsilon_{res} = \varepsilon - \frac{Na}{\sqrt{2}L}$$  Eq. 12

It is now possible to evaluate the free enthalpy of formation of nucleates accounting for the effect of misfit dislocations.

The Fig. 8 (a) shows both the interfacial contribution (black dotted line) and the elastic contribution (black dashed line) with no dislocations (as in Fig. 6). The elastic energy with $N$ dislocations per facet, $\Delta G'_{elas} V + E_{dislo}$, (from black to grey when increasing $N$) has also been plotted as a function of the precipitate size. The number of dislocations per facet ($N$) that provides an elastic contribution lower than the interfacial contribution is shown to depend on the precipitate size ($L$), see Table 2. The optimal number of dislocations minimizing the most the elastic contribution obviously increases with the precipitate size. Besides there is always a configuration providing an elastic contribution lower than the interfacial one. This theoretical approach fully explain the (111) tetrahedral shaped precipitates. If we extract the minimum value of the elastic energy as a function of size, we obtain the black dashed line in Fig. 8 (b). This curve is not smooth because the curvature jumps each time a new dislocation is introduced. The corresponding change in Gibbs energy $\Delta G$ is also represented in black in Fig. 8 (b). This energy is now becoming negative for a certain size and exhibits a maximum corresponding to the critical size ($L^* = 1.7$ nm). By introducing an appropriate number of misfit dislocations, precipitate can nucleate and grow.

It is worth mentioning that the same calculations for a platelet-shaped precipitates lead to the same conclusions. In contrast, nucleation of cube shaped precipitate with (100) facets is not possible. If cuboidal precipitates with (100) facets would have nucleated, the above-mentioned soft directions in CuAg indicate that growth of those precipitates would have been controlled by the elastic stress. However, without introducing any dislocation, the work of formation of such precipitate is increasing with its size and never becomes negative (infinite barrier). Besides, when introducing dislocations, the interfacial contribution to the work of formation is too large to explain the presence of (100) facets. Therefore, cuboidal nuclei cannot nucleate.

Considering thermodynamics we have explained the physical reasons for nucleating tetrahedron shaped precipitates in the binary Cu-Ag system. However two kinds of precipitation modes were observed experimentally after 30' at 440°C: precipitation in the grains that may be either homogeneous or heterogeneous (on lattice defects) and discontinuous precipitation. Another ageing temperature (2h at 400°C) and another ageing time at the same temperature (15' at 440°C) have also been experimentally investigated. They exhibit both transformation modes of the supersaturated solid solutions as well. However the percentage of the analyzed material transformed by one mode or another is different in every case. It is then interesting to combine some thermodynamic and kinetics aspects to understand these observations.

**4.3. The occurrence of continuous precipitation**

After the different ageing times and treatments, both discontinuous and continuous transformations were observed. However the amount of discontinuous precipitates compared to inner-grains precipitates varies as well are their size and number density.

For shorter ageing times (15' at 440°C, against 30' described before), the number density of continuous precipitates was found higher and their size two to three times smaller. This indicates that, from a kinetic point of view, the coarsening stage has already been reached. Nevertheless, the APT chemical analysis revealed a concentration in the matrix and a precipitate volume fraction in grains continuously transformed still far from equilibrium. Besides, the amount of grains discontinuously transformed is higher after 30' than after 15' at 440°C. This indicates that the initially continuously transformed grains may be captured by the discontinuous front. It was actually observed [14] that the growth of continuous precipitates in Cu-Ag-Zr requires the formation and migration of ledges along the interface, a rate-limiting mechanism which may reduce the growth rate. Consequently the transformation of the still supersaturated solid solution is easier with the discontinuous mode. After 2h at 400°C the proportion of discontinuously transformed zones is much more important than after 15 or 30' at 440°C. Thus the influence of the temperature on the occurrence of the continuous transformation needs to be clarified.

The driving force for nucleation, either homogeneous or heterogeneous, and both the nucleation barrier and the critical size vary with the temperature.

Fig. 9 (a) and (b) displays, for two temperatures T=600°C and T=200°C with dashed lines grey and black respectively, the elastic contribution as a function of size when introducing misfit dislocations and extracting the lowest elastic contribution. Indeed the latter varies with configuration and size. A procedure similar to that explained before for Fig. 8 has been applied. The interfacial contribution and the work of formation are represented with black dotted line and black solid line respectively. The elastic contribution to the Gibbs energy of the system is reduced when introducing dislocations. Furthermore the interfacial term becomes also greater than the elastic one and control the shape of precipitates. Again, the work of formation of a nucleus can be decreased and becomes favorable when introducing dislocations.

Series of calculations were made for different temperatures between 200 and 600°C. Homogeneous nucleation was shown to be always possible when introducing dislocations. Both the nucleation barrier and the critical size decrease when the temperature decreases (Fig. 8 (b) and Fig. 9) as expected. However, this does not explain why less continuously transformed precipitates have been observed at 400°C. Furthermore the typical diffusion distance of Ag in Cu during 2h at 400°C is relatively close to that at 440°C during 30'. Indeed, an extrapolation of an Arrhenius law fitting experimental data [26] gives a bulk diffusion distance for Ag in fcc Cu of 70 and 60 nm respectively in both cases.

Then the relative importance of precipitation modes, i.e. in the grain and discontinuous, should be considered. The thermodynamics of grain boundaries nucleation has to be addressed because it is the first step of discontinuous precipitation. The barrier for this process can be calculated using the model developed by Clemm and Fisher [27]. In their approach, they described the work of formation of a heterogeneous precipitate considering the balance between the former grain boundary interface α/α' with an interfacial energy $\gamma_{AA}$ and the new precipitate/matrix interface of a certain surface with another interfacial energy $\gamma_{AB}$. No elastic contribution is taken into account because the interface is assumed to be fully incoherent. The barrier W* is then expressed by:

$$W^* = \frac{4}{27} \frac{(b\gamma_{AB} - a\gamma_{AA})^2}{c^2 \Delta G_{ch}^2}$$ Eq. 13

with $a = \pi(1-k^2)$, $b = 4\pi(1-k)$, $c = \frac{2\pi}{3}(2-3k+k^2)$ and $k = \frac{\gamma_{AA}}{2\gamma_{AB}}$. $\gamma_{AA}$ is given by the value of Cu-Cu grain boundary energy (0.8 J.m$^{-2}$[28]). $\gamma_{AB} = \gamma_{111}$ given in Table 1.

Both homogeneous and heterogeneous nucleation barriers are plotted as a function of temperature in Fig. 10. The heterogeneous nucleation barrier is the smallest for the full temperature range. It is important to note that the influence of the interfacial energy has been checked (data not shown here) and that the trend is unchanged in the range of 0.5 < γ < 1 J.m$^{-2}$. The ratio between the homogeneous nucleation barrier and the heterogeneous nucleation barrier is shown in Fig. 10 (b) as a function of temperature. The evolution as a function of the temperature is clearly not monotonous. This ratio is found to be the smallest near 450°C, close to the ageing temperature used in the present study where abundant homogeneous precipitation was observed in the grains. Contrariwise, at 400°C, this ratio is higher. From a physical point of view, it means that the misfit dislocations and their appropriate number at the critical size do not have the same weight for accommodating the lattice mismatch and preserving a thermodynamic stability depending on the temperature, i.e. depending on the available driving force for the transformation. A balance has to be found and it works better for some temperatures.

The thermodynamics of the two precipitation modes suggests that the discontinuous mode is always favored. It is then important to understand why this mode is not exclusive at the two temperatures mentioned in this study. First, it is worth mentioning that the available nucleation sites for the discontinuous mechanism, i.e. grain boundaries, are in a much lower volume density than sites for homogeneous nucleation in the matrix. Therefore homogeneous nucleation may be present in the material although the barrier is larger, as experimentally observed in the present work (Fig. 3, Fig. 4 and Fig. 5). On top of that, if nucleation within grains takes place on defects, the nucleation barrier can be minimized. The situation can then be reversed and the inner-grain nucleation may become dominant.

The defects potentially assisting the inner-grain nucleation may be vacancies or vacancy clusters [29]. Once nucleated, vacancy clusters grow during annealing and their number density remain constant. Furthermore, it has been shown for example that in quenched and aged gold the vacancy precipitation led to the formation of stacking-fault tetrahedrons [29]. A similar phenomenon could also occur in Cu-5%Ag because of the supersaturation of vacancies induced by quenching prior to the precipitation treatment. Once the vacancy clusters created, Ag solute atoms can diffuse towards these defects.

Vacancy supersaturation may be evaluated at equilibrium [30]:

$$X_V = K_{V0} \exp\left(\frac{-E_{VF}}{kT}\right) \qquad \text{Eq. 14}$$

where $K_{V0}$ is a constant related to entropy and $E_{VF}$ is the formation energy of a vacancy. Diffusion coefficient expresses as:

$$D_V = D_{V0} \exp\left(\frac{-Q}{kT}\right) \qquad \text{Eq. 15}$$

where $Q$ is the activation energy for the vacancy diffusion.

Table 3 summarizes the values of parameters in Eq. 14 and Eq. 15 and compile values of $X_V$ and $D_V$ for various temperatures.

At 800°C, the initial vacancy concentration is about $5 \times 10^{-6}$ while the precipitate number density was estimated to be in a range of $10^{21}$ to $10^{22}$ m$^{-3}$ from our experimental data. Let us assume that all quenched vacancies (from 800°C) form vacancy tetrahedrons in the early stages of annealing. If the number density is taken equal to the observed precipitate number density, then the size of tetrahedral precipitates should be in a range of 1.5 to 3.5 nm. This is close to the critical size for nucleation estimated previously at 440°C. Heterogeneous nucleation on vacancy clusters may thus significantly promote nucleation. This nucleation mechanism may also explain the non-equilibrium value of the matrix concentration obtained by APT. Indeed the number of available nucleation sites is limited by the vacancy supersaturation. When they are all occupied, nucleation is difficult. On the early stages of growth of precipitates, the parent phase remains supersaturated. Thus the easiest way to reduce the supersaturation may be through the movement of the DP front.

During isothermal ageing (2h), the vacancies are fast enough to diffuse within a grain in the temperature range [200°C-600°C]. However at higher temperatures, the difference between the equilibrium vacancy concentration at the ageing temperature and at the homogenization temperature is much smaller. The vacancy supersaturation is likely to be too small to promote the formation of vacancy clusters acting as nucleation sites.

Moreover, on a kinetic aspect, at low temperature, even if the nucleation barrier is lower, Ag diffusion is so slow that discontinuous precipitation dominates due to fast transport in the GB.

All these considerations highlight the complexity of mechanisms involved during the transformation of the supersaturated solid solution. It is worth mentioning that to understand the competition between continuous and discontinuous precipitation, the study of the mobility of the grain boundaries, a key parameter in discontinuous transformations that depends on the misorientation, is needed. This was not taken into account in this work.

## 5. Conclusions

Continuously transformed precipitates in Cu-5%wtAg aged 30' at 440°C have been investigated. The peculiar shape of inner-grains precipitates observed with both APT and STEM analysis is due to the anisotropy of the interfacial energy. The elastic energy is significantly decreased by misfit dislocations at the interface between the precipitate and the matrix that in turns decreases the nucleation barrier and makes the (111) oriented, tetrahedral and platelet-shaped precipitates thermodynamically favorable. Both thermodynamic and kinetic aspects have to be taken into account to understand the occurrence of this continuous transformation against the discontinuous that is thermodynamically more favorable. However, close to the ageing temperature of the study the ratio between the homogeneous nucleation barrier and the heterogeneous one is minimized as compared to other temperatures. In combination with a small number of DP nucleation sites (namely grain boundaries), it leads to a significant decomposition of the solid solution through homogeneous precipitation. This work was more especially dedicated to the study of homogeneous nucleation in Cu-Ag. Nevertheless a complementary study on discontinuous precipitation is of a great importance to fully catch the temperature and composition dependence in the occurrence of the different precipitation mechanisms.

# References

[1] Highconductivity Copper for electrical engineering,… - Google Scholar, (n.d.). https://scholar.google.fr/scholar?hl=fr&as_sdt=0%2C5&q=Highconductivity+Copper+for+electrical+engineering%2C+Copper+development+association%2C+CDA+publication+122%2C+UK%2C+1998.+&btnG= (accessed February 13, 2018).

[2] P.R. Subramanian, J.H. Perepezko, The ag-cu (silver-copper) system, J. Phase Equilibria. 14 (1993) 62–75. doi:10.1007/BF02652162.

[3] Y. Champion, J. Bourgon, X. Sauvage, Modified strain rate regime in ultrafine grained copper with silver micro-alloying, Mater. Sci. Eng. A. 657 (2016) 1–5. doi:10.1016/j.msea.2016.01.044.

[4] S.M. Foiles, M.I. Baskes, M.S. Daw, Embedded-atom-method functions for the fcc metals Cu, Ag, Au, Ni, Pd, Pt, and their alloys, Phys. Rev. B. 33 (1986) 7983–7991. doi:10.1103/PhysRevB.33.7983.

[5] Y. Sakai, T. Hibaru, K. Miura, A. Matsuo, K. Kawaguchi, K. Kindo, Development of High Strength-High Conductivity Cu-6 wt% Ag Alloy for High Field Magnet, MRS Adv. 1 (2016) 1137–1148. doi:10.1557/adv.2015.4.

[6] F. Bittner, S. Yin, A. Kauffmann, J. Freudenberger, H. Klauß, G. Korpala, R. Kawalla, W. Schillinger, L. Schultz, Dynamic recrystallisation and precipitation behaviour of high strength and highly conducting Cu–Ag–Zr-alloys, Mater. Sci. Eng. A. 597 (2014) 139–147. doi:10.1016/j.msea.2013.12.051.

[7] C. Zhao, X. Zuo, E. Wang, R. Niu, K. Han, Simultaneously increasing strength and electrical conductivity in nanostructured Cu–Ag composite, Mater. Sci. Eng. A. 652 (2016) 296–304. doi:10.1016/j.msea.2015.11.067.

[8] J.L. Murray, Calculations of Stable and Metastable Equilibrium Diagrams of the Ag-Cu and Cd-Zn Systems, Metall. Trans. A. 15 (1984) 261–268. doi:10.1007/BF02645110.

[9] K.N. Tu, D. Turnbull, Morphology of cellular precipitation of tin from lead-tin bicrystals, Acta Metall. 15 (1967) 369–376. doi:10.1016/0001-6160(67)90214-3.

[10] R.A. Fournelle, J.B. Clark, The genesis of the cellular precipitation reaction, Metall. Trans. 3 (1972) 2757–2767. doi:10.1007/BF02652842.

[11] P. Bacher, P. Wynblatt, S.M. Foiles, A Monte Carlo study of the structur and composition of (001) semicoherent interphase boundaries in Cu–Ag–Au alloys, Acta Metall. Mater. 39 (1991) 2681–2691. doi:10.1016/0956-7151(91)90084-E.

[12] D. Raabe, P.-P. Choi, Y. Li, A. Kostka, X. Sauvage, F. Lecouturier, K. Hono, R. Kirchheim, R. Pippan, D. Embury, Metallic composites processed via extreme deformation: Toward the limits of strength in bulk materials, MRS Bull. 35 (2010) 982–991. doi:10.1557/mrs2010.703.

[13] A. Gaganov, J. Freudenberger, E. Botcharova, L. Schultz, Effect of Zr additions on the microstructure, and the mechanical and electrical properties of Cu–7 wt.%Ag alloys, Mater. Sci. Eng. A. 437 (2006) 313–322. doi:10.1016/j.msea.2006.07.121.

[14] W. Piyawit, W.Z. Xu, S.N. Mathaudhu, J. Freudenberger, J.M. Rigsbee, Y.T. Zhu, Nucleation and growth mechanism of Ag precipitates in a CuAgZr alloy, Mater. Sci. Eng. A. 610 (2014) 85–90. doi:10.1016/j.msea.2014.05.023.

[15] R. Räty, H.M. Miekk-Oja, Precipitation associated with the growth of stacking faults in copper–silver alloys, Philos. Mag. J. Theor. Exp. Appl. Phys. 18 (1968) 1105–1125. doi:10.1080/14786436808227742.

[16] M. Bonvalet, T. Philippe, X. Sauvage, D. Blavette, The influence of size on the composition of nano-precipitates in coherent precipitation, Philos. Mag. 94 (2014) 2956–2966. doi:10.1080/14786435.2014.941029.

[17] W. Lefebvre, F. Vurpillot, X. Sauvage, Atom Probe Tomography: Put Theory Into Practice, Academic Press, 2016.

[18] J.-O. Andersson, T. Helander, L. Höglund, P. Shi, B. Sundman, Thermo-Calc & DICTRA, computational tools for materials science, Calphad. 26 (2002) 273–312. doi:10.1016/S0364-5916(02)00037-8.

[19] Thermo-Calc Software AB, TCCU2: Cu-based Alloys Database, 2017. http://www.thermocalc.com/media/41191/tccu2.pdf (accessed February 13, 2018).

[20] A. Devaraj, D.E. Perea, J. Liu, L.M. Gordon, T.J. Prosa, P. Parikh, D.R. Diercks, S. Meher, R.P. Kolli, Y.S. Meng, S. Thevuthasan, Three-dimensional nanoscale characterisation of materials by atom probe tomography, Int. Mater. Rev. 63 (2018) 68–101. doi:10.1080/09506608.2016.1270728.

[21] J. Roesler, H. Harders, M. Baeker, Mechanical Behaviour of Engineering Materials: Metals, Ceramics, Polymers, and Composites, Springer Science & Business Media, 2007.

[22] J.D. Eshelby, The Continuum Theory of Lattice Defects, in: F. Seitz, D. Turnbull (Eds.), Solid State Phys., Academic Press, 1956: pp. 79–144. doi:10.1016/S0081-1947(08)60132-0.

[23] J.W. Cahn, On spinodal decomposition in cubic crystals, Acta Metall. 10 (1962) 179–183. doi:10.1016/0001-6160(62)90114-1.

[24] R.C. Lincoln, K.M. Koliwad, P.B. Ghate, Morse-Potential Evaluation of Second- and Third-Order Elastic Constants of Some Cubic Metals, Phys. Rev. 157 (1967) 463–466. doi:10.1103/PhysRev.157.463.

[25] J.P. Hirth, J. Lothe, Theory of Dislocations, McGraw-Hill, 1968.

[26] D.B. Butrymowicz, J.R. Manning, E.R. Read, Diffusion in Cu and Cu alloys, J. Phys. Chem. 5 (1976). http://www.nist.gov/data/PDFfiles/jpcrd76.pdf (accessed January 19, 2016).

[27] P.J. Clemm, J.C. Fisher, The influence of grain boundaries on the nucleation of secondary phases, Acta Metall. 3 (1955) 70–73. doi:10.1016/0001-6160(55)90014-6.

[28] B. Runnels, I.J. Beyerlein, S. Conti, M. Ortiz, An analytical model of interfacial energy based on a lattice-matching interatomic energy, J. Mech. Phys. Solids. 89 (2016) 174–193. doi:10.1016/j.jmps.2016.01.008.

[29] K.C. Jain, R.W. Siegel, On the growth of annealing of stacking-fault tetrahedra in gold, Philos. Mag. J. Theor. Exp. Appl. Phys. 26 (1972) 637–647. doi:10.1080/14786437208230110.

[30] L. Karlsson, H. Nordén, H. Odelius, Overview no. 63 Non-equilibrium grain boundary segregation of boron in austenitic stainless steel—I. Large scale segregation behaviour, Acta Metall. 36 (1988) 1–12. doi:10.1016/0001-6160(88)90023-5.

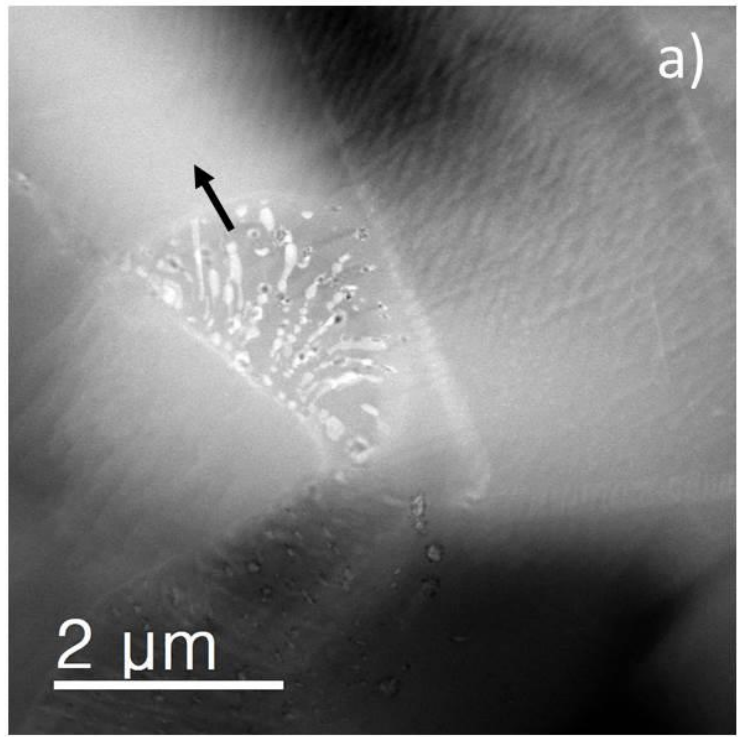

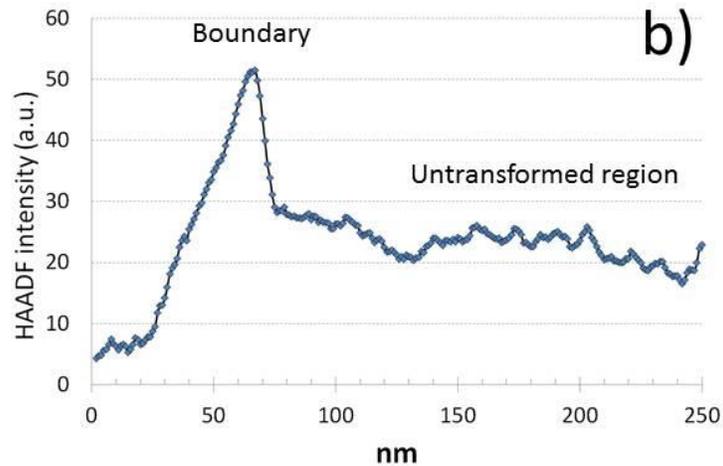

Figure 1: (a) STEM HAADF image showing a grain that has started to transform via discontinuous precipitation and containing typical lamella shape Ag particles aligned along the growth direction (brightly imaged), (b) HAADF intensity line profile computed from the HAADF image across the transformation front (direction arrowed on (a)) showing a higher intensity at the interface related to an enhanced Ag content at the transformation front.

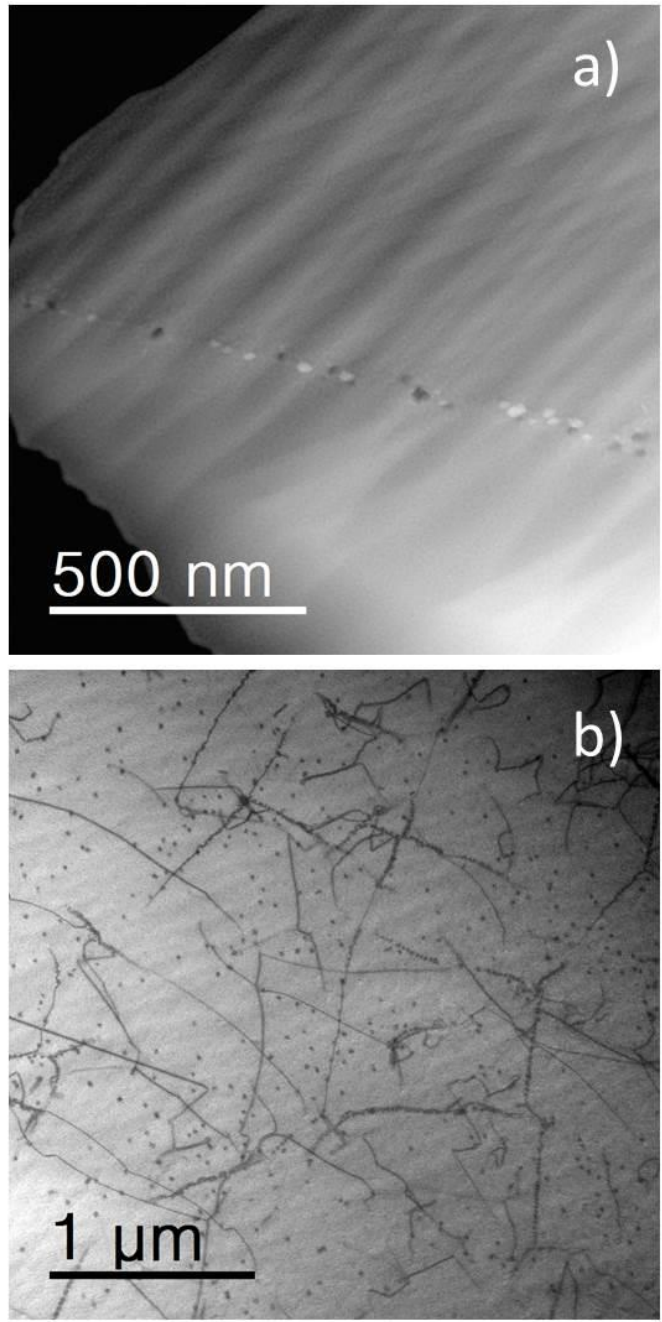

Figure 2: (a) STEM HAADF image showing brightly imaged Ag precipitates that have nucleated at a GB (dark patches are Ag particles lost during the sample preparation), (b) STEM BF images showing Ag precipitates(dark spots) that have homogeneously precipitated in the matrix and heterogenously precipitated along dislocations (dark lines).

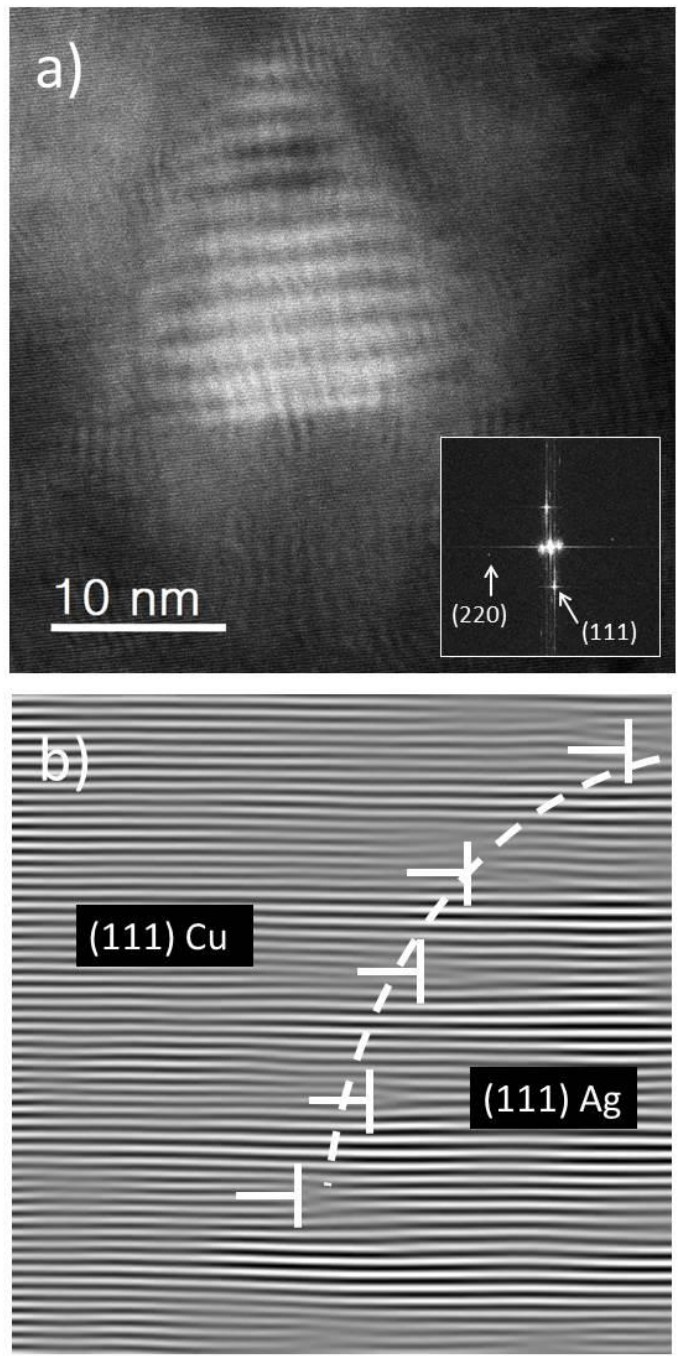

Figure 3: (a) High resolution STEM HAADF image of a nanoscaled Ag precipitate viewed as a triangle in (112) zone axis (inset: indexed FFT), (b) Fourier filtered image showing (111) edge dislocations that relax the misfit between (111) fcc Cu and Ag planes.

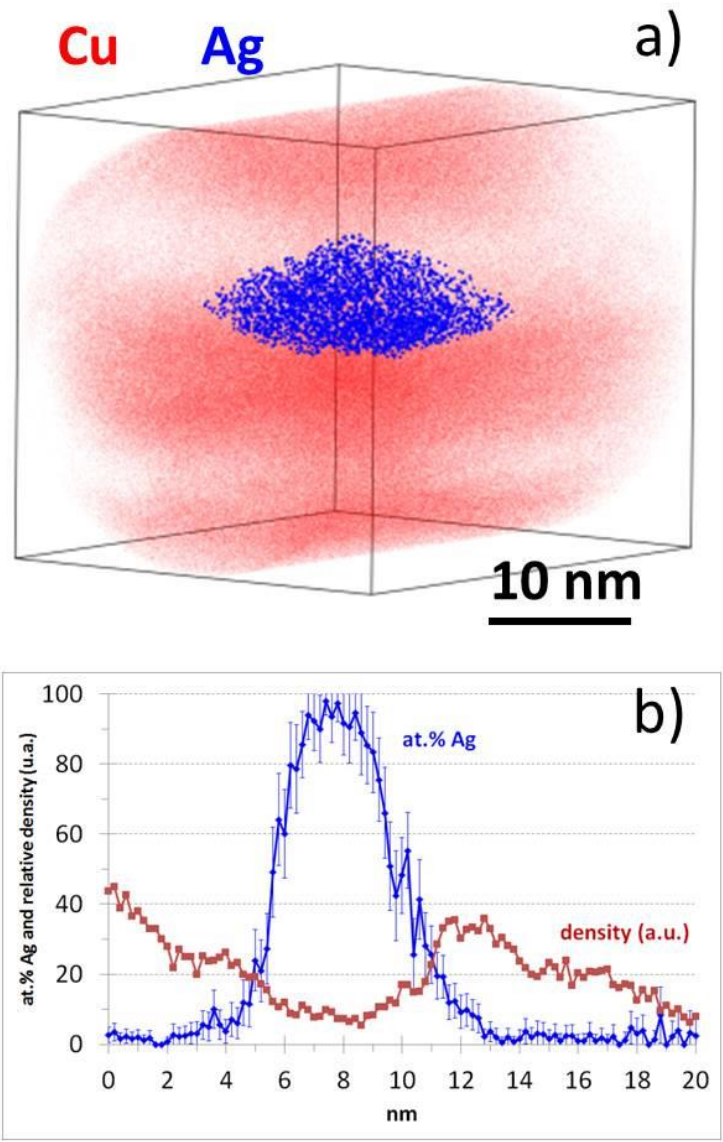

Figure 4: (a) Three dimensional reconstruction of a volume analyzed by APT (30x30x35 nm$^3$) showing a silver precipitate with a tetrahedron shape (filtered data where Ag is shown only where the Ag concentration exceeds 20at.%, Ag atoms are plotted in blue and Cu in red), (b) Silver concentration profile computed across the precipitate (sampling volume thickness 1nm) and local density showing fluctuations related to local magnification effects.

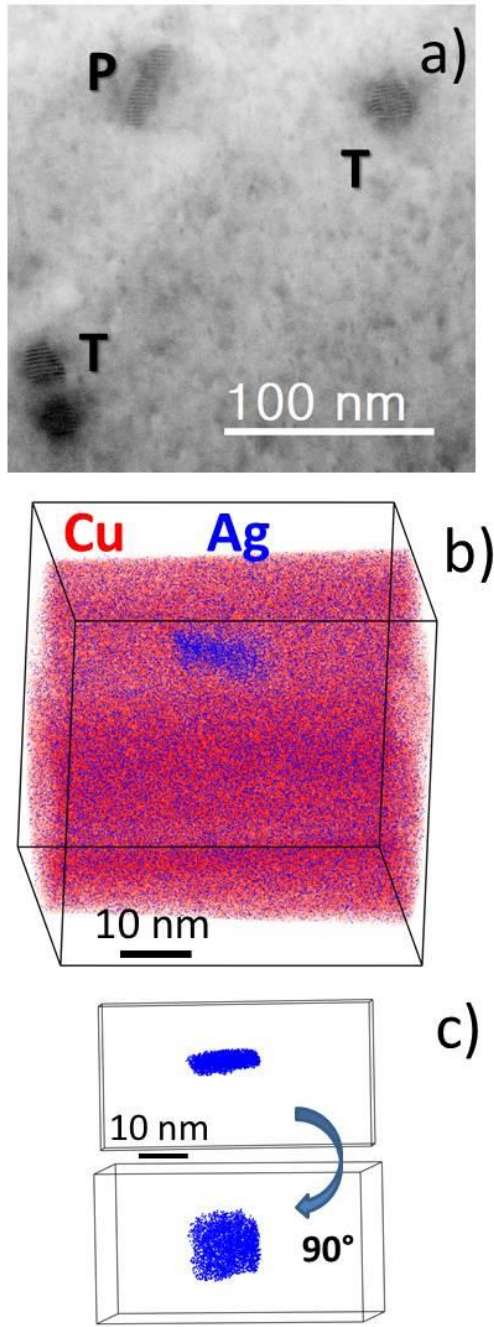

Figure 5: (a) STEM-BF image showing Ag tetrahedrons (labelled T) and an Ag plate (labelled P), (b) Three dimensional reconstruction of a volume analyzed by APT (50x50x50 nm$^3$) showing a silver particle with a plate shape (Ag blue and Cu red), (c) the same particle viewed in two perpendicular directions.

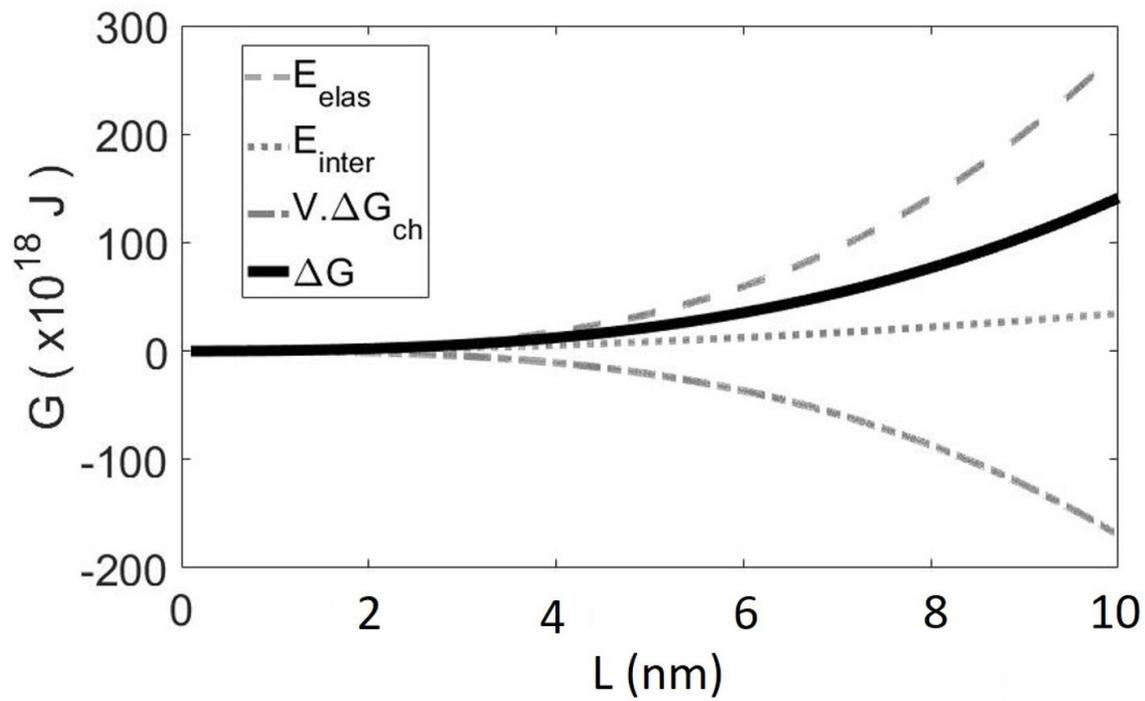

Figure 6: Work of formation of a tetrahedral nucleus of size *L* (black solid line). This energy encompasses in three terms: the elastic contribution (dashed line), the interfacial contribution (dotted line) and the chemical contribution (dashed-dotted line).

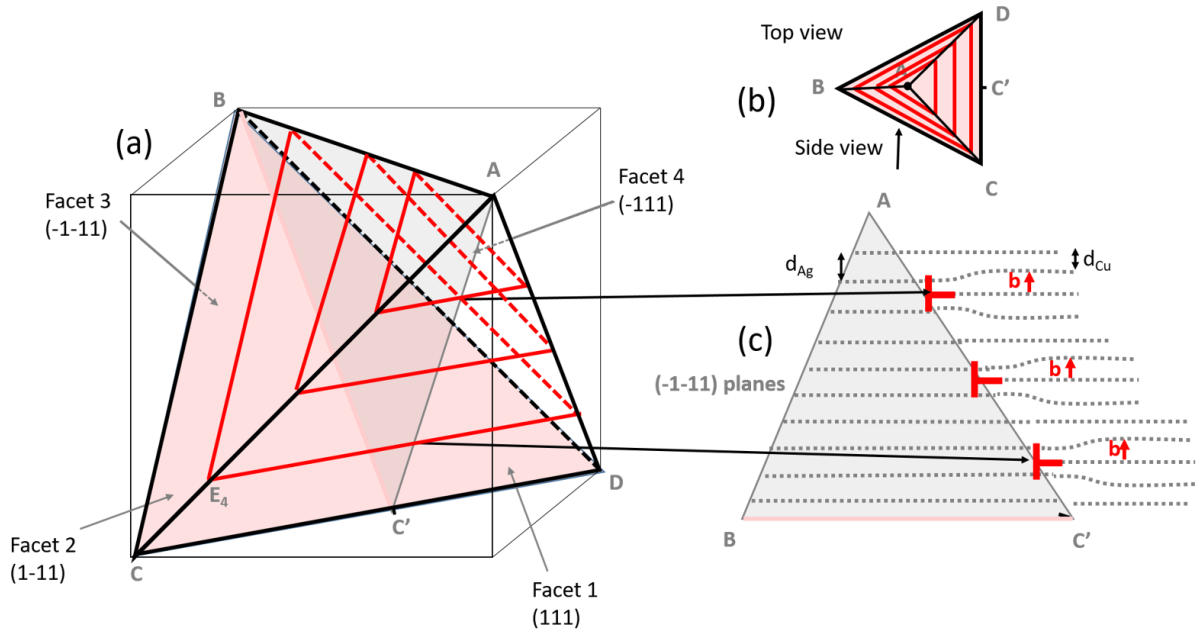

Figure 7: (a) Sketch of dislocation circuits running around three facets of the tetrahedral precipitate for N=3 and with a 1/3 <-1-11> Burgers' vector. C' is the middle of the CD edge. (b) Top view of the same tetrahedral precipitate depicting the same three dislocation circuits. (c) Side view of the same tetrahedral precipitate. The slice corresponds to the shaded triangle in (a). The grey dotted lines depict the (111) planes and show the relaxation of the lattice thanks to the dislocations with a 1/3 <-111> Burgers' vector.

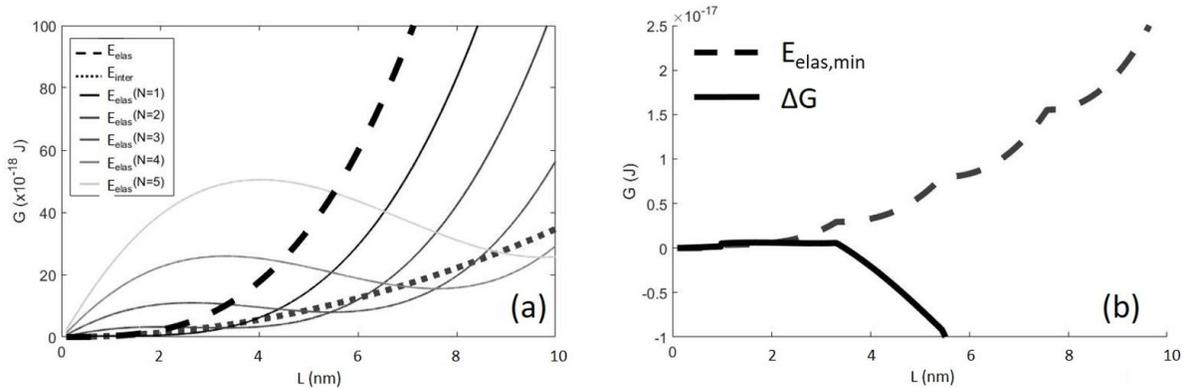

Figure 8: Interfacial (black dotted line) and elastic without any dislocation (black dashed line) contribution to the work of formation of a nucleus. When introducing dislocations the elastic contribution is modified (black to light grey when increasing $N$ – (b) Selection of the minimum elastic contribution (the dashed line follows the line made by the lowest values of the black to light grey curves in (a)) and resulting work of formation (black solid line) of a nucleus. The critical size is equal to 1.7 nm.

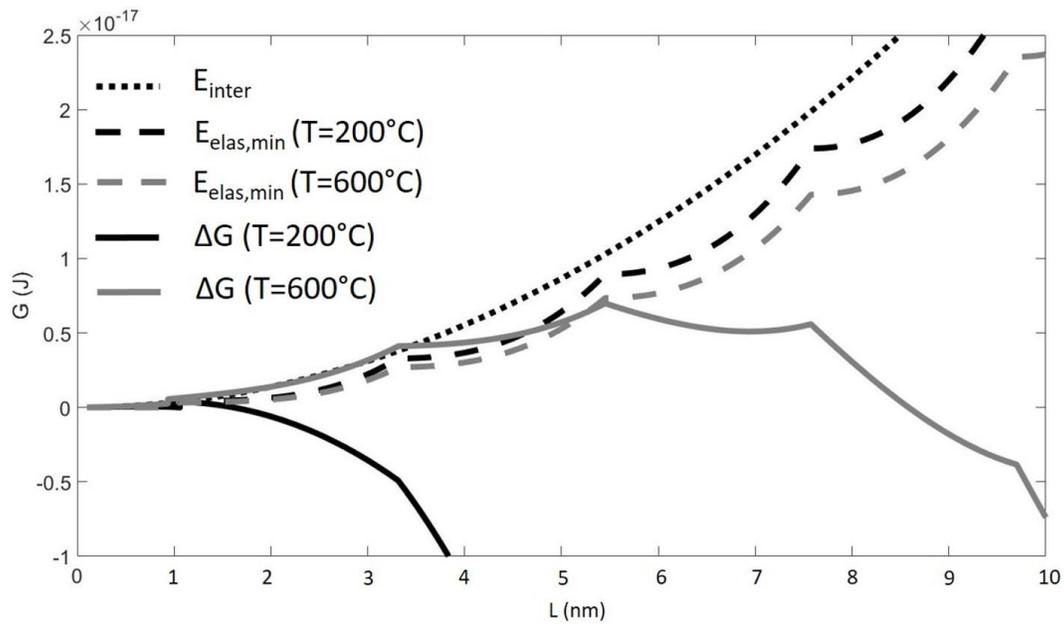

Figure 9: Interfacial (black dotted line) contribution to the work of formation of a nucleus at T=200°C and T=600°C. When introducing dislocations the elastic contribution (dashed lines) and work of formation (solid lines) are modified at both T=200°C (black) and T=600°C (grey). The procedure for the selection of the minimum elastic contribution as a function of size is the same than for obtaining Fig. 8(b). The critical sizes are equal to 5.5 nm and 1 nm for T=600°C and T=200°C respectively.

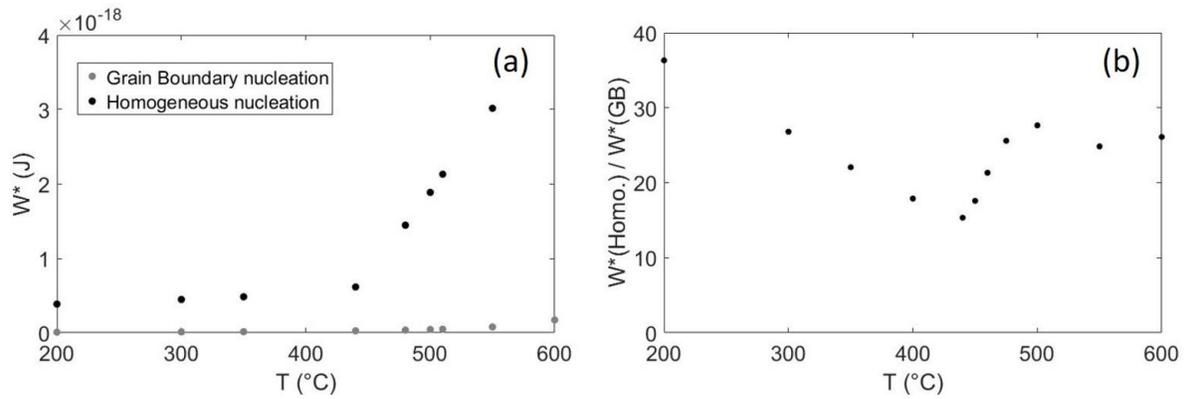

Figure 10: (a) Nucleation barrier as a function of temperature for homogeneous (grey dots) and heterogeneous nucleation at grain boundary (GB) (black dots) – (b) Ratio between homogeneous nucleation barrier and the GB nucleation barrier as a function of temperature